\documentclass[oldversion,letter]{aa} 
\usepackage{graphicx}
\usepackage{txfonts}

\usepackage{natbib}
\bibpunct{(}{)}{;}{a}{}{,} 

\begin{document}
   \title{The physical origin of the X-ray power spectral density break timescale in accreting black holes}

   \subtitle{}

   \author{W. Ishibashi
          \inst{1}
          \and
          T. J.-L. Courvoisier
          \inst{2,3}
          }

   \institute{ Institute of Astronomy, Madingley Road, Cambridge CB3 0HA, UK 
      \and ISDC Data Centre for Astrophysics, ch. d'Ecogia 16, 1290 Versoix, Switzerland 
    \and Geneva Observatory, Geneva University, ch. des Maillettes 51, 1290 Sauverny, Switzerland \\
    \email{wako@ast.cam.ac.uk, Thierry.Courvoisier@unige.ch} }

\date{Received; accepted}


   \abstract{X-ray variability of active galactic nuclei (AGN) and black hole binaries can be analysed by means of the power spectral density (PSD). The break observed in the power spectrum defines a characteristic variability timescale of the accreting system. The empirical variability scaling that relates characteristic timescale, black hole mass, and accretion rate ($T_B \propto M_{BH}^{2.1}/\dot{M}^{0.98}$) extends from supermassive black holes in AGN down to stellar-mass black holes in binary systems. We suggest that the PSD break timescale is associated with the cooling timescale of electrons in the Comptonisation process at the origin of the observed hard X-ray emission. We find that the Compton cooling timescale directly leads to the observational scaling, which naturally reproduces the functional dependence on black hole mass and accretion rate ($t_C \propto M_{BH}^{2}/\dot{M}$). This result simply arises from general properties of the emission mechanism and is independent of the details of any specific accretion model. }



   \authorrunning{ W. Ishibashi \and T. J.-L. Courvoisier}
   \titlerunning{X-ray PSD break timescale in accreting black holes}

   \maketitle
%

\section{Introduction}

X-ray variability is a characteristic property of all accreting black hole systems, including supermassive black holes in active galactic nuclei (AGN) and stellar-mass black holes in binary systems (black hole binaries, BHB). 
The observation of rapid and aperiodic variations indicates that X-ray emission is produced in the innermost regions of the accretion flow, close to the central black hole. 
The study of X-ray variability is thus thought to provide information on the physical emission mechanisms and possibly on the nature of the underlying accretion flows. 

Quantitative analysis of X-ray variability is often performed by means of the power spectral density (PSD). 
The  observed variability power spectrum can be roughly modelled by power laws of the form $P_{\nu} \propto \nu^{\alpha}$, where $\nu = 1/T$ is the temporal frequency and $\alpha$ the power law slope.  
A typical AGN power spectrum is characterised by a slope of $\alpha \sim -2$ at high frequencies (corresponding to short timescales), flattening to a slope of $\alpha \sim -1$ at lower frequencies (or equivalently longer timescales).
A similar shape is observed in the power spectra of galactic black holes in the high/soft state. 
In addition, a second break to a slope of $\alpha \sim 0$, occurring at still lower frequencies, is seen in the low/hard state of black hole binaries. In the following we consider the high-frequency PSD break, $\nu_B$ (from $\alpha \sim -1$ to $\alpha \sim -2$), in both AGN and black hole binaries. 
The corresponding break timescale, $T_B = 1/\nu_B$, is considered to be a characteristic X-ray variability timescale of the accreting system. 
Typical break timescales measured in AGN are of $\sim$day-timescales, while the corresponding break timescales seen in black hole binaries are of the order of fractions of seconds \citep[][]{McH_LNP, Gilfanov_LNP}. 
The difference in break timescales observed between supermassive and stellar-mass black holes is broadly consistent with a simple scaling with black hole mass, and linear mass-timescale relations have been proposed \citep{Markowitz_et_2003, Papadakis_2004}.  

A more quantitative X-ray variability scaling has been obtained by \citet{McHardy_et_2006}, who considered the explicit  dependence of the characteristic timescale on both black hole mass and accretion rate. 
A major result is that this observational scaling relation extends from supermassive black holes in AGN down to stellar-mass black holes in binary systems, covering many orders of magnitude in black hole mass.
The fact that both AGN and black hole binaries seem to follow the same empirical scaling has been claimed to be a strong support for the scale invariance of accreting black hole systems, and led to the suggestion that AGN are simply scaled-up galactic black holes \citep{McHardy_et_2006}. 
This in turn led to the proposal of the so-called `variability fundamental plane', which relates characteristic timescale, black hole mass, and accretion rate.  

However, the physical origin of the PSD break is not definitively established yet. 
Various interpretations of the PSD break timescale have been proposed in the literature, trying to link the observed break timescales with different accretion model timescales. 
At first, the measured break timescales have been compared with characteristic physical timescales of the accretion disc, including orbital, thermal, and viscous timescales at different radii \citep{Markowitz_et_2003, Papadakis_2004}. 
More recently, the PSD break has been discussed in the framework of the inner propagating accretion flow fluctuations model, in which  fluctuations originating at different outer radii propagate inwards, leading to modulations of the accretion rate at inner radii and hence to variations in the observed X-ray flux \citep{Lyubarskij_1997, Churazov_et_2001}. 
The optically thick accretion disc is assumed to be truncated at some distance from the centre and the inner region is replaced by a hot, optically thin flow component. In this picture, the break timescale is associated with a characteristic timescale, usually the viscous time, at the truncation radius of the standard accretion disc. 
In an alternative approach, we have discussed the characteristic X-ray variability timescale within the framework of a clumpy accretion model in AGN \citep{Paper_2}.
In this scenario, accretion onto the central black hole occurs through a sequence or cascade of shocks, which are at the origin of the observed radiation.  
Accreting elements first interact in the form of optically thick shocks, giving rise to optical/UV emission, later followed by optically thin shocks giving rise to X-rays through inverse Compton emission. 
Optical/UV photons emitted in the optically thick shocks form the seed photons for Compton-cooling of the electrons heated in the optically thin shocks. 
The characteristic heating/cooling timescale of the electrons in the optically thin shocks has been associated with the PSD break timescale.

Although the exact geometry of the inner coronal flow is still unknown, hard X-ray emission is widely attributed to Compton processes.  
In the inverse Compton process, soft seed photons (possibly emitted as thermal emission from the accretion disc) are upscattered to higher energies by energetic electrons in a hot, optically thin corona. 
The observed hard X-ray power law component of accreting black holes is indeed well represented by Comptonisation models \citep[][and references therein]{Titarchuk_1994}. 
As hard X-rays are produced by the Comptonisation mechanism, it seems natural to consider physical timescales associated with the emission process itself as possible candidates for the identification with the observed X-ray PSD break timescale. 
In the following we consider the characteristic timescale associated with the Compton cooling of the hot electrons. 


\section{Compton cooling timescale}
\label{Sect_Compton_cooling}

The power emitted by a single electron in the inverse Compton scattering is given by \citep{R_L_1979}
\begin{equation}
L_C = \frac{4}{3} c \sigma_T \gamma^2 \beta^2 u_{\gamma} \, , 
\label{L_Compton}
\end{equation}
where $\sigma_T$ is the Thomson cross section and $u_{\gamma}$ the photon energy density. 

The Compton cooling timescale, which sets the characteristic timescale for the cooling of electrons by inverse Compton emission, is defined as
\begin{equation}
t_C = \frac{E_e}{L_C} \, , 
\label{eq_tCdef}
\end{equation}
where $E_e$ is the average electron energy. 
In the non-relativistic limit ($\gamma \sim 1$, $\beta^2 \propto E_e$), the dependence on the electron energy cancels, and 
Eq. (\ref{eq_tCdef}) leads to
\begin{equation}
t_C = \frac{3 m_e c}{8 \sigma_T} \frac{1}{u_{\gamma}} \, . 
\label{eq_tNR}
\end{equation}

The radiation density at distance $R$ is given by
\begin{equation}
u_{\gamma} = \frac{L_s}{4 \pi R^2 c} \, , 
\end{equation}
where $L_s$ is the seed photon luminosity for inverse Compton scattering. 
Assuming that the seed photons are generated in the radiatively efficient accretion process, the seed photon luminosity is directly proportional to the accretion rate, $L_s = \eta \dot{M} c^2$, where $\eta$$\sim$0.1 is the canonical radiative efficiency. 
The radial distance $R$ can be expressed in units of the Schwarzschild radius as $R = \zeta R_S$, where $R_S = \frac{2 G M_{BH}}{c^2}$. 

The Compton cooling timescale (Eq. \ref{eq_tNR}) is now written as
\begin{equation}
t_C = \frac{6 \pi G^2 m_e}{\sigma_T c^4} \frac{\zeta^2}{\eta} \frac{M_{BH}^2}{\dot{M}}  \, . 
\label{t_C}
\end{equation} 

We see that the Compton cooling time is mainly determined by two of the black hole fundamental parameters, i.e. mass and accretion rate. 
As the Compton cooling time is directly linked to the physical process of X-ray emission, it can be considered to be a characteristic X-ray timescale of the accreting system.  

For  a typical AGN of mass $M_{BH} \sim 10^7 M_{\odot}$ and accreting in the radiatively efficient regime (with typical Eddington ratios of $L/L_E \sim 0.1$), the numerical value of the characteristic timescale (Eq. \ref{t_C}) is of the order of

\begin{equation}
t_{AGN} \sim 0.1 \, \eta_{0.1}^{-1} \zeta_{100}^{2} \dot{M}_{-2}^{-1} M_7^{2} \; \mathrm{d} \, , 
\label{t_AGN}
\end{equation}
where $M_{BH} = M_7 \cdot 10^7 M_{\odot}$ and $\dot{M} = \dot{M}_{-2} \cdot 0.01 M_{\odot}$/yr. 
We see that the characteristic Compton cooling timescale in AGN is of $\sim$day-timescales.  
We can also estimate the Compton cooling timescale in black hole binaries by introducing parameters appropriate for galactic black holes in Eq. (\ref{t_C}). Considering an object of about $M_{BH} \sim10M_{\odot}$ accreting at a few percent of the Eddington limit, and taking into account the fact that the seed photon luminosity is only a fraction of the bolometric luminosity, the characteristic timescale is of the order of $t_{BHB} \sim 0.07$s.

We observe that the Compton cooling timescale is shorter for small black hole mass and/or high accretion rate, scaling as
\begin{equation}
t_C \propto \frac{M_{BH}^2}{\dot{M}} \, . 
\end{equation}

The Compton cooling of the hot electrons leads to X-ray emission. 
The average electron energy can be estimated by assuming an equilibrium situation between Compton cooling and Coulomb heating rates: $L_C = L_H$ \citep{Paper_1}. The resulting electron energy scales as $E_e \propto n_e^{2/7} u_{\gamma}^{-2/7}$, where $n_e$ is the electron number density. 
The average X-ray luminosity is given by the Compton luminosity of a single electron (Eq. \ref{L_Compton}) multiplied by the average number of electrons:
\begin{equation}
\langle L_X \rangle \sim L_C \cdot N_e \propto n_e^{9/7} u_{\gamma}^{5/7} \, , 
\end{equation}
and we see that the dependence on the electron number density is stronger than that on the photon energy density.  
Thus X-ray luminosity variations are mainly driven by electron density modulations associated with the inhomogeneities of the upscattering medium. 
In the clumpy accretion scenario, the origin of the density modulations was associated with the inhomogeneous structure of the accretion flow resulting from the succession of shocks \citep{Paper_2}. 
Here we note that the physical origin of the modulations can be diverse and does not need to be specified. 
Density fluctuations on timescales shorter than the electron cooling time ($t_{f} < t_C$), which at equilibrium is equal to the electron heating time, cannot be propagated and are smoothed out. The suppression of fluctuations on timescales shorter than a critical value, given by the Compton cooling time, induces a break in the observed variability power spectrum, which may be associated with the PSD break.


\section{Comparison with the PSD break timescale}

Observational measurements of the high-frequency PSD break (from $\alpha \sim -1$ to $\alpha \sim -2$) have been performed for a number of AGN and galactic black holes. 
The observed break timescales in Seyfert galaxies are usually in the range $\sim$0.01 to a few days \citep{Markowitz_et_2003, McH_et_2004, U_McH_2005}, while the corresponding break timescales measured in black hole binaries are of the order of seconds or less \citep[][and references therein]{Gilfanov_LNP}. 
The Compton cooling timescales estimated in the previous section ($\sim$day-timescales in AGN and fractions of seconds in BHB) thus correspond to the typically measured values of the PSD break in both AGN and galactic black holes. 
The characteristic timescale therefore seems to scale roughly linearly with black hole mass, from stellar-mass black holes up to supermassive black holes. 
However, an additional dependence on the accretion rate is required by more detailed observations of AGN: for a given black hole mass, higher accretion rate objects have shorter break timescales than lower accretion rate counterparts \citep{U_McH_2005}.
The empirical relationship between variability timescale, black hole mass, and accretion rate has been obtained by \citet{McHardy_et_2006} based on a main sample of AGN including two galactic black holes in radio-quiet states. The resulting empirical scaling can be written as

\begin{equation}
T_B \sim \frac{M_{BH}^{1.12}}{\dot{m}_E^{0.98}} \, , 
\label{eq_obs}
\end{equation}
where $\dot{m}_E = L_{bol}/L_E$ is the ratio of the bolometric luminosity to Eddington luminosity. 
This observational scaling is shown to extend from AGN down to stellar-mass black holes in binary systems. 
As the Eddington luminosity is given by the black hole mass ($L_E \propto M_{BH}$) and assuming that the bolometric luminosity is proportional to the accretion rate ($L_{bol} \propto \dot{M}$), the empirical scaling (Eq. \ref{eq_obs}) can be re-expressed directly in terms of the physical parameters as
\begin{equation}
T_B \sim \frac{M_{BH}^{2.1}}{\dot{M}^{0.98}} \, .
\label{T_B_obs}
\end{equation}

Comparison of the Compton cooling timescale, $t_C \propto \frac{M_{BH}^2}{\dot{M}}$ (derived in Eq. \ref{t_C}), with the observational scaling (Eq. \ref{T_B_obs}) shows a very close match. We observe that the characteristic timescale associated with the Compton cooling process directly reproduces the observed dependence on black hole mass and accretion rate. 
We note that a similar result has been previously derived but within the specific framework of the clumpy accretion model in AGN \citep{Paper_2}. 


\section{Spectral-timing relation}

X-ray variability properties are also closely coupled with X-ray spectral properties. 
Observations of `spectral-timing' correlations have been reported in both AGN and black hole binaries, whereby an increase in the characteristic frequency is correlated with a steepening of the X-ray spectrum. 
The characteristic noise frequency is observed to increase with the steepening of the energy spectrum in Cygnus X-1 \citep{Gilfanov_et_1999}, and similar patterns are also observed in other X-ray binary systems, e.g. GX 339-4 \citep{Revnivtsev_et_2001}. 
In the case of AGN, the average X-ray spectral slope is found to be positively correlated with the normalized characteristic frequency defined as $\nu_{norm} = \nu_B \times M_{BH}$, where $\nu_B$ is the break frequency observed in the X-ray PSD \citep{Papadakis_et_2009}. 

It is known that Comptonisation processes lead to power law spectra, and the resulting X-ray spectra observed in the hard energy band are usually described by power laws, characterised by the photon index $\Gamma$. 
A significant correlation between photon index and Eddington ratio has been reported in numerous AGN samples \citep{Shemmer_et_2008, Kelly_et_2008, Risaliti_et_2009}, suggesting that the X-ray spectral slope is correlated with some form of accretion rate. 
Within the Comptonisation scenario, the physical interpretation of the observed $\Gamma - L/L_E$ correlation is that an increase in the accretion rate leads to an enhancement in the seed photon emission, which in turn leads to efficient Compton cooling of the hot corona, which finally leads to a steepening of the hard X-ray spectrum.  
On the other hand, we have seen that the Compton cooling time scales inversely with the seed photon energy density (Eq. \ref{eq_tNR}), thus an increase in the seed photon emission also leads to a decrease in the characteristic timescale.
Combining the two results we may account for the observed correlation between characteristic frequency and spectral steepening. 
Therefore in the context of Comptonisation, both the X-ray characteristic timescale and the spectral slope seem to be determined by the supply of seed photons, which in turn depends on the accretion rate.  
Hence, for a given black hole mass, higher accretion rate systems tend to have both steep spectra and short characteristic timescales, in agreement with the observed `spectral-timing correlation' \citep{Papadakis_et_2009}.  
This suggests that both timing and spectral properties of accreting black hole systems are mainly determined by the accretion rate parameter.


\section{Discussion}

The break observed in the X-ray PSD is thought to provide insight into the underlying variability mechanism and various physical interpretations of the break timescale have been proposed. 
As a first approach, the observed PSD break timescales have been compared with physical timescales of the accretion disc in the innermost regions of the accretion flow where X-ray emission is thought to be produced. 
The shortest timescale is given by the dynamical or orbital timescale, $t_{\phi}$. 
This is of the order of $\sim$hours for typical AGN parameters and thus too short compared to the break timescales of the order of $\sim$days observed in Seyfert galaxies. On the other hand, for a standard geometrically thin disc (with aspect ratio $H/R << 1$) the viscous timescales (of the order of  $\gtrsim$years) are found to be a few orders of magnitude longer compared to the observed values \citep{Markowitz_et_2003}.   
The closest values to the measured break timescales are perhaps given by thermal timescales, and thermal instabilities have also been considered \citep{Papadakis_2004}. 
We note that the physical timescales of the accretion disc are all proportional to the black hole mass and thus the approximate scaling of the characteristic timescale with black hole mass is easily obtained; however, the dependence on the accretion rate is less obvious and cannot be simply accounted for. 

The currently favoured interpretation of the PSD break is based on the inner propagating fluctuations model \citep{Lyubarskij_1997, Churazov_et_2001}. The variability power is found to decrease on short timescales, with a rapid fall-off below the break, suggesting that the break timescale might be associated with some cut-off in the accreting system. 
Accordingly, the break timescale has been associated with the viscous timescale ($t_{\nu} \sim t_{\phi}/ \alpha_{ss} \, (H/R)^2$) at the inner edge of the optically thick accretion disc. 
The accretion disc is assumed to be truncated at some distance from the centre, the location of the truncation radius being connected with the accretion rate: at high accretion rates the optically thick disc extends down to the last stable orbit, while at low accretion rates the disc is truncated at large distances \citep[][and references therein]{Zdziarski_Gierlinski_2004}. 
In this picture, the value of the characteristic timescale is related to the position of the disc truncation radius. 
If the disc is truncated far away from the centre, the associated viscous timescale is long; while with increasing accretion rate, the truncation radius becomes smaller and the associated characteristic timescale correspondingly shorter. 
Once the accretion disc has reached the last stable orbit, a further decrease of the viscous timescale can only be obtained by varying the disc thickness parameter (H/R), assuming that the H/R ratio increases with increasing accretion rate  \citep{McH_LNP}. 
Therefore by combining variations of the disc truncation radius with variations of the disc thickness parameter one may qualitatively account for the observed trend with accretion rate. 

Although the PSD break timescale has usually been associated with some characteristic timescale of the accretion disc, it is well known that the accretion disc itself cannot account for the hard X-ray emission component of accreting black holes. 
Consequently, the identification of a characteristic disc timescale at a particular radius with the characteristic X-ray variability timescale is not straightforward, and other interpretations may be considered. 
On the other hand, it is widely agreed that hard X-ray emission is produced by Comptonisation processes. 
It seems therefore interesting to see whether there is any characteristic timescale associated with the emission process itself that might be compared with the observed PSD break timescale. 

We have previously associated the PSD break timescale with the heating/cooling time of the electrons in the optically thin shocks within the clumpy accretion scenario \citep{Paper_2}. 
Here we suggest that the X-ray PSD break timescale could be associated with the cooling time of electrons in the Comptonisation process, independently of the details of any specific accretion model. 
The characteristic X-ray variability timescale is simply given by the Compton cooling time;  the observed scaling with black hole mass and accretion rate is quantitatively reproduced, without additional assumptions. 
It is quite remarkable that the observational scaling simply derives from the definition of the Compton cooling timescale with a basic assumption of a radiatively efficient accretion and a quasi-spherical distribution of the seed photon population. 

The Comptonisation process is thought to operate in a similar way in supermassive and stellar-mass black holes \citep{Gliozzi_et_2011}.
We have seen that numerical estimates of the Compton cooling time give correct order-of-magnitude values of the PSD break timescales in AGN and binary systems.
In our picture, the characteristic timescale is set by the Compton cooling time in both AGN and galactic black holes, suggesting that the underlying emission mechanism is scale-invariant. 
In addition, if the characteristic X-ray variability timescale is given by the Compton cooling timescale, this may additionally support the hypothesis that the main mechanism of hard X-ray emission is indeed Comptonisation in both classes of accreting black holes.


\section{Conclusion}

Various physical interpretations of the PSD break have been proposed in the literature. 
The observed break timescale is usually associated with some characteristic timescale of the accretion disc, such as the viscous timescale at the truncation radius. 
Here we suggest another possibility and associate the break timescale with a physical timescale directly linked to the X-ray emission mechanism, i.e. the cooling time of the electrons in the Comptonisation process.
The association seems natural since hard X-rays are generated in the Comptonisation process and it also provides a possible physical interpretation of the PSD break.
We obtain that the Compton cooling time directly reproduces the correct functional dependence on black hole mass and accretion rate, defining the empirical variability scaling of \citet{McHardy_et_2006}. 
We show that this result is model-independent while the detailed physical realisations of the accretion flows might take different forms in different accreting systems.


\begin{acknowledgements}
 W.I. acknowledges support from the Swiss National Science Foundation.     
\end{acknowledgements}

\bibliographystyle{aa}
\bibliography{biblio.bib}

\begin{thebibliography}{21}
\expandafter\ifx\csname natexlab\endcsname\relax\def\natexlab#1{#1}\fi

\bibitem[{{Churazov} {et~al.}(2001){Churazov}, {Gilfanov}, \&
  {Revnivtsev}}]{Churazov_et_2001}
{Churazov}, E., {Gilfanov}, M., \& {Revnivtsev}, M. 2001, \mnras, 321, 759

\bibitem[{{Gilfanov}(2010)}]{Gilfanov_LNP}
{Gilfanov}, M. 2010, in Lecture Notes in Physics, Berlin Springer Verlag, Vol.
  794, Lecture Notes in Physics, Berlin Springer Verlag, ed. {T.~Belloni}, 17

\bibitem[{{Gilfanov} {et~al.}(1999){Gilfanov}, {Churazov}, \&
  {Revnivtsev}}]{Gilfanov_et_1999}
{Gilfanov}, M., {Churazov}, E., \& {Revnivtsev}, M. 1999, \aap, 352, 182

\bibitem[{{Gliozzi} {et~al.}(2011){Gliozzi}, {Titarchuk}, {Satyapal}, {Price},
  \& {Jang}}]{Gliozzi_et_2011}
{Gliozzi}, M., {Titarchuk}, L., {Satyapal}, S., {Price}, D., \& {Jang}, I.
  2011, \apj, 735, 16

\bibitem[{{Ishibashi} \& {Courvoisier}(2009{\natexlab{a}})}]{Paper_1}
{Ishibashi}, W. \& {Courvoisier}, T.~J.-L. 2009{\natexlab{a}}, \aap, 495, 113

\bibitem[{{Ishibashi} \& {Courvoisier}(2009{\natexlab{b}})}]{Paper_2}
{Ishibashi}, W. \& {Courvoisier}, T.~J.-L. 2009{\natexlab{b}}, \aap, 504, 61

\bibitem[{{Kelly} {et~al.}(2008){Kelly}, {Bechtold}, {Trump}, {Vestergaard}, \&
  {Siemiginowska}}]{Kelly_et_2008}
{Kelly}, B.~C., {Bechtold}, J., {Trump}, J.~R., {Vestergaard}, M., \&
  {Siemiginowska}, A. 2008, \apjs, 176, 355

\bibitem[{{Lyubarskii}(1997)}]{Lyubarskij_1997}
{Lyubarskii}, Y.~E. 1997, \mnras, 292, 679

\bibitem[{{Markowitz} {et~al.}(2003){Markowitz}, {Edelson}, {Vaughan},
  {Uttley}, {George}, {Griffiths}, {Kaspi}, {Lawrence}, {McHardy}, {Nandra},
  {Pounds}, {Reeves}, {Schurch}, \& {Warwick}}]{Markowitz_et_2003}
{Markowitz}, A., {Edelson}, R., {Vaughan}, S., {et~al.} 2003, \apj, 593, 96

\bibitem[{{McHardy}(2010)}]{McH_LNP}
{McHardy}, I. 2010, in Lecture Notes in Physics, Berlin Springer Verlag, Vol.
  794, Lecture Notes in Physics, Berlin Springer Verlag, ed. {T.~Belloni}, 203

\bibitem[{{McHardy} {et~al.}(2006){McHardy}, {Koerding}, {Knigge}, {Uttley}, \&
  {Fender}}]{McHardy_et_2006}
{McHardy}, I.~M., {Koerding}, E., {Knigge}, C., {Uttley}, P., \& {Fender},
  R.~P. 2006, \nat, 444, 730

\bibitem[{{McHardy} {et~al.}(2004){McHardy}, {Papadakis}, {Uttley}, {Page}, \&
  {Mason}}]{McH_et_2004}
{McHardy}, I.~M., {Papadakis}, I.~E., {Uttley}, P., {Page}, M.~J., \& {Mason},
  K.~O. 2004, \mnras, 348, 783

\bibitem[{{Papadakis}(2004)}]{Papadakis_2004}
{Papadakis}, I.~E. 2004, \mnras, 348, 207

\bibitem[{{Papadakis} {et~al.}(2009){Papadakis}, {Sobolewska}, {Arevalo},
  {Markowitz}, {McHardy}, {Miller}, {Reeves}, \& {Turner}}]{Papadakis_et_2009}
{Papadakis}, I.~E., {Sobolewska}, M., {Arevalo}, P., {et~al.} 2009, \aap, 494,
  905

\bibitem[{{Revnivtsev} {et~al.}(2001){Revnivtsev}, {Gilfanov}, \&
  {Churazov}}]{Revnivtsev_et_2001}
{Revnivtsev}, M., {Gilfanov}, M., \& {Churazov}, E. 2001, \aap, 380, 520

\bibitem[{{Risaliti} {et~al.}(2009){Risaliti}, {Young}, \&
  {Elvis}}]{Risaliti_et_2009}
{Risaliti}, G., {Young}, M., \& {Elvis}, M. 2009, \apjl, 700, L6

\bibitem[{{Rybicki} \& {Lightman}(1979)}]{R_L_1979}
{Rybicki}, G.~B. \& {Lightman}, A.~P. 1979, {Radiative processes in
  astrophysics}, ed. {Rybicki, G.~B.~\& Lightman, A.~P.}

\bibitem[{{Shemmer} {et~al.}(2008){Shemmer}, {Brandt}, {Netzer}, {Maiolino}, \&
  {Kaspi}}]{Shemmer_et_2008}
{Shemmer}, O., {Brandt}, W.~N., {Netzer}, H., {Maiolino}, R., \& {Kaspi}, S.
  2008, \apj, 682, 81

\bibitem[{{Titarchuk}(1994)}]{Titarchuk_1994}
{Titarchuk}, L. 1994, \apj, 434, 570

\bibitem[{{Uttley} \& {McHardy}(2005)}]{U_McH_2005}
{Uttley}, P. \& {McHardy}, I.~M. 2005, \mnras, 363, 586

\bibitem[{{Zdziarski} \& {Gierli{\'n}ski}(2004)}]{Zdziarski_Gierlinski_2004}
{Zdziarski}, A.~A. \& {Gierli{\'n}ski}, M. 2004, Progress of Theoretical
  Physics Supplement, 155, 99

\end{thebibliography}

\end{document}